\RequirePackage{fixltx2e}
\documentclass[5p,twocolumn]{elsarticle}

\usepackage{bm}
\usepackage{graphicx}
\usepackage{amsbsy}
\usepackage{amstext}
\usepackage{amssymb}
\usepackage{float}
\usepackage{hyperref}

\AtBeginDocument{}

\begin{document}

\title{Magnetic oscillations in silicene}

\author[uns,ifisur]{F.~Escudero\corref{cor1}}
\ead{federico.escudero@uns.edu.ar}
\author[uns,ifisur]{J.S.~Ardenghi}
\ead{jsardenhi@gmail.com}
\author[uns,ifisur]{P.~Jasen}
\ead{pvjasen@uns.edu.ar}

\cortext[cor1]{Corresponding author}
\address[uns]{Departamento de F{\'i}sica, Universidad Nacional del Sur, \\
	Av. Alem 1253, B8000CPB, Bah{\'i}a Blanca, Argentina}

\address[ifisur]{Instituto de F{\'i}sica del Sur (IFISUR, UNS-CONICET), \\
	Av. Alem 1253, B8000CPB, Bah{\'i}a Blanca, Argentina}

\begin{abstract}
In this work the magnetic oscillations (MO) in pristine silicene at $T=0$
K are studied. Considering a constant electron density we obtain analytical expressions
for the ground state internal energy and magnetization, under a perpendicular
electric and magnetic field, taking in consideration the Zeeman effect.
It is found that the MO are sawtooth-like, depending on the change in the last occupied energy level. This leads us to a classification
of the MO peaks in terms of the Landau level (LL), valley or spin changes. Using this
classification we analyze the MO for different values of the electric
field $E_{z}$. When $E_{z}=0$, the energy levels have a valley degeneracy
and the MO peaks occur only whenever the last energy level changes
its LL and/or spin. When $E_{z}\neq0$, the valley degeneracy is broken
and new MO peaks appear, associated with the valley change in the
last energy level. By analyzing the MO peaks amplitude it is possible to extract information about the Fermi velocity and the spin-orbit
interaction strength. Finally we analyze the MO frequencies, which
can also be associated with the change of LL, valley or spin in the
last energy level.
\end{abstract}

\maketitle

\section{Introduction}

In the past few years silicene has been gaining considerable interest
in the scientific community \cite{Kara_2012,Houssa_2015,Zhuang_2015}. Like graphene,
silicene has a 2D hexagonal structure with silicon atoms at each lattice
site, with two interpenetrating sublattices \emph{A} and \emph{B}.
The reciprocal space is also a hexagonal lattice in the momentum space,
which in turn defines the Brillouin zone. Silicene is best described
with a tight binding (TB) model, which leads to an effective Dirac-like
Hamiltonian in the low energy approximation, with the sublattices
\emph{A} and \emph{B} acting as a pseudospin degree of freedom \cite{Guzm_n_Verri_2007,Houssa_2011,Cahangirov_2009}.
Nevertheless, silicene distinguish itself from graphene by two important
features. One is the large spin-orbit interaction (SOI), about 3.9
meV \cite{Liu_2011} (compared to $10^{-3}$ meV in graphene \cite{Yao_2007}),
which makes silicene a topological insulator. Moreover, this strong
SOI would make possible the observation of the quantum spin Hall effect \cite{Liu2011,An_2013,Ezawa_2012,Ezawa_2013}. The other characteristic is that the lattice
structure in silicene is not planar but buckled, with a layer separation
between the two sublattices \cite{Houssa_2015}. Thus by introducing a
potential difference between the two sublattices one can tune the
bandgap \cite{Drummond_2012,Ni_2012,Liu_2014,Vargiamidis_2014}. These features imply
than in silicene at low energies the electrons behave as massive Dirac
fermions \cite{2016}, moving with a Fermi velocity of about $5.5\times10^{5}$
m/s \cite{Guzm_n_Verri_2007,Dzade_2010}.

When a magnetic field is applied to silicene, the discrete Landau
levels (LL) are obtained. As in graphene, due to the relativistic-like
dispersion relation these levels are not equidistant \cite{Tabert_2013}.
Moreover, the Landau energy in silicene is smaller than in graphene,
due to the bigger Fermi velocity in the latter. The LL create
an oscillating behavior in the thermodynamics potentials. For instance,
the magnetization oscillates as a function of the inverse magnetic
field, the so called de Haas van Alphen (dHvA) effect \cite{Shoenberg_1952}.
The different frequencies involved in the oscillations are related
to the closed orbits that electrons perform on the Fermi surface.
This effect is purely quantum mechanical and is an useful tool to
map the Fermi surface \cite{Onsager_1952}. In graphene it has been found
that, without impurities, the magnetization oscillates periodically
in a sawtooth pattern \cite{Sharapov_2004,Zhang_2010}. It is then expected
that in silicene the magnetization also oscillates in a sawtooth pattern. 

Because of the buckled nature of silicene, by applying a perpendicular
electric field the spin and valley degeneracy of the LL is lifted \cite{Shakouri_2014}. In this
case, in contrast with graphene, the energy levels of each valley
are different and there is no more valley degeneracy. Moreover, considering
the Zeeman effect, the LL for each spin split, loosing then the spin
degeneracy. This loss of valley and spin degeneracy gives discontinuous
changes in the last energy level, which in turn is expected to produce
new magnetization peaks, as occur in graphene \cite{Escudero_2017}. Motivated
by this we have studied the magnetic oscillations (MO) at $T=0$ K
in a general silicene-like system with a conduction electron density
$n_{e}$, which could be due to an applied gate voltage. 

We have organized
the work as follow: In Sec. 2 we obtain the energy levels of
silicene in a perpendicular magnetic and electric field, considering
the intrinsic SOI and the Zeeman effect. From this we obtain
an expression for the ground state internal energy
and magnetization. 
In Sec. 3 we study and classify the MO peaks for different values of perpendicular electric field. 
In Sec. 4 we analyze the MO frequencies by performing a fast Fourier transform (FFT). Finally our conclusions follow in Sec. 5.

\section{Magnetic oscillations in silicene }

\subsection{Energy spectrum}

In the low wavelength approximation, with energies near the Fermi
energy, the electrons in silicene are described by a Dirac-like Hamiltonian
in 2D for massive fermions. In a perpendicular electric field $E_{z}$
it reads \cite{2016}

\begin{equation}
H_{\eta s}=\upsilon_{F}(\eta\sigma_{x}p_{x}+\sigma_{y}p_{y})+\sigma_{z}\Delta_{\eta s},\label{H free}
\end{equation}
where $\upsilon_{F}\sim5.5\times10^{5}$ m/s is the Fermi velocity \cite{Dzade_2010}, $\sigma_{i}$ are the Pauli matrices acting in the
sublattices \emph{A} and \emph{B}, $\eta=1\:(-1)$ for the valley
$K$ ($K'$), $s=\pm1$ for spin and down, and 
\begin{equation}
\Delta_{\eta s}=\eta s\lambda_{SO}-elE_{z},
\end{equation}
where $\lambda_{SO}$ is the intrinsic spin-orbit interaction (SOI)
strength and $l$ is the buckle height. We shall consider a perpendicular
magnetic field $B$, so that $\mathbf{B}=B\mathbf{e}_{z}$. In the
Landau gauge we have $\mathbf{A}=-By\mathbf{e}_{x},$ and the momentum
changes following the Peierls substitution \cite{Peierls_1933} $\mathbf{p}\rightarrow\mathbf{p}-e\mathbf{A}$.
Considering the Zeeman effect \cite{Zeeman_1897}, the term $\bm{\mu}\cdot\mathbf{B}=\mu_{B}gBs_{z}/2$
is added to $H$, where $s_{z}=2S_{z}/\hbar$ is the Pauli matrix
acting in the spin state. We omit the nearest- and next-nearest neighbor
Rashba SOIs, denoted as $\lambda_{R1}$ and $\lambda_{R2}$ in \cite{Ezawa_2012},
since they are negligible in comparison to the intrinsic SOI $\lambda_{SO}$.
Then Eq. (\ref{H free}) becomes

\begin{equation}
H_{\eta s}=\upsilon_{F}\left[\eta\sigma_{x}\left(p_{x}+eBy\right)+\sigma_{y}p_{y}\right]+\sigma_{z}\Delta_{\eta s}-\mathbf{\boldsymbol{\mu}}\cdot\mathbf{B}.\label{H field}
\end{equation}
Because $H$ only depends on the $y$ coordinate, we can express the
wave function as $\psi=e^{-ikx}(\begin{array}{cc}
\psi^{A} & \psi^{B}\end{array})$, with $\psi^{A/B}$ depending only on $y$. Then, introducing the
ladder matrices $\sigma_{\pm}=\sigma_{x}\pm i\sigma_{y}$ and making
the change of variable \cite{Ardenghi_2013} $y'=\left(-\hbar k+eBy\right)/\sqrt{\hbar eB}$,
the equation $H\psi=E\psi$ becomes

\begin{eqnarray}
	\biggl\{\upsilon_{F}\sqrt{\hbar eB}\biggl[\frac{\sigma_{+}}{2}\left(\eta y'-\partial_{y'}\right) & + & \frac{\sigma_{-}}{2}\left(\eta y'+\partial_{y'}\right)\biggr]\nonumber\\
	+\sigma_{z}\Delta_{\eta s}-\mathbf{\boldsymbol{\mu}}\cdot\mathbf{B}\biggr\}\psi_{\eta s} & = & E\psi_{\eta s}.
\end{eqnarray}
Defining the ladder operators $a^{\dagger}=\left(y'-\partial_{y'}\right)/\sqrt{2}$
and $a=\left(y'+\partial_{y'}\right)/\sqrt{2}$ we get 

\begin{eqnarray}
\biggl[\eta\frac{\hbar\omega_{L}}{2}\left(\sigma_{+}\alpha_{\eta}^{\dagger}+\sigma_{-}\alpha_{\eta}\right) & + & \sigma_{z}\Delta_{\eta s}\nonumber\\
-\hbar\omega_{Z}s_{z}\biggr]\psi_{\eta s} & = & E\psi_{\eta s},\label{H-E}
\end{eqnarray}
where $\alpha_{1}=a$, $\alpha_{-1}=a^{\dagger}$ and $\omega_{L}=\upsilon_{F}\sqrt{2eB/\hbar}$,
$\omega_{Z}=\mu_{B}gB/2\hbar$.
The energies can be calculated by writing the wave function for each
valley and spin as

\begin{equation}
	\left|\psi_{s}^{\eta}\right\rangle =b_{s}^{\eta}\left|n,A,s\right\rangle +c_{s}^{\eta}\left|n-\eta,B,s\right\rangle ,
\end{equation}
where $b_{s}^{\eta}$ and $c_{s}^{\eta}$ are constants, $n$ is the
Landau level (LL) index and $\left|s\right\rangle =\left|\pm\right\rangle $
represents the spin state, so that $s_{z}\left|s\right\rangle =s\left|s\right\rangle $.
Then, given that $\sigma_{+}\left|A\right\rangle =0$, $\sigma_{+}\left|B\right\rangle =2\left|A\right\rangle $,
$\sigma_{-}\left|A\right\rangle =2\left|B\right\rangle $, $\sigma_{-}\left|B\right\rangle =0$,
$\sigma_{z}\left|A\right\rangle =\left|A\right\rangle $, $\sigma_{z}\left|B\right\rangle =-\left|B\right\rangle $
and $a^{\dagger}\left|n\right\rangle =\sqrt{n+1}\left|n+1\right\rangle $,
$a\left|n\right\rangle =\sqrt{n}\left|n-1\right\rangle $, solving
Eq. (\ref{H-E}) the energy spectrum results

\begin{eqnarray}
E_{0,\eta,s} & = & s\lambda_{SO}-\eta elE_{z}-s\hbar\omega_{Z}\quad\left(n=0\right),\label{energy n=0}\\
E_{n,\eta,s,\beta} & = & \beta\sqrt{\left(s\lambda_{SO}-\eta elE_{z}\right)^{2}+\left(\hbar\omega_{L}\right)^{2}n}\label{energies}\nonumber\\
&  & -s\hbar\omega_{Z}\quad\left(n\geq1\right),
\end{eqnarray}
where $\beta=-1$ for the valence band (VB) and $\beta=1$ for the conduction band
(CB). When $E_{z}=0$, the LL have a doubly valley degeneracy, whereas
when $E_{z}\neq0$ this degeneracy vanishes. Notice that without
the Zeeman effect the LL $n=0$ is always twice less degenerate than
the LL $n\geq1$, regardless of $E_{z}$ \cite{Shakouri_2014}. Therefore the
Zeeman effect gives an equal degeneracy for all LL. Moreover, as in
the classical case, each LL has a degeneracy due to the free
direction ($x$ in this case) which is not quantized. This degeneracy
comes by imposing periodical boundary conditions and is given by $D=AB/\varphi$,
where $A$ is the silicene sheet area and $\varphi=h/e$ is the magnetic
unit flux.

\subsection{Ground state magnetization}

We shall study the ground state magnetization ($T=0$ K) for this
system, under the influence of a perpendicular magnetic and electric
field, where the energy levels are given by Eqs. (\ref{energy n=0}) and (\ref{energies}).
We consider a constant electron density $n_{e}=N/A$, which may due
to an applied gate voltage, such that valence band is full and only
the conduction band is available. The valence band would still make a continuous (non-oscillatory)
contribution to the magnetization, but since we are interested only
in the MO, we will not take this contribution into account. The internal energy at
$T=0$ K for the $N$ conduction electrons can be computed as the
sum of the filled Landau levels. The number of totally filled levels
is $q=\left[q_{c}\right],$ where $q_{c}=N/D$ is the filling factor,
and the brackets means the biggest integer less or equal to $q_{c}$
(Floor function). It is worth noting that we assume that
$N$ is constant, instead of the the chemical potential $\mu$ (Fermi
energy) being constant. However, for both cases the results are similar (see the Appendix
for details).

In order to compute the ground state internal energy we have to sort the energy levels. We call $\xi_{m}$ the decreasing
sorted energy levels, $m=0,1,2\ldots$ being the label index. In general
we write\footnote{For the LL $n=0$ only the positive root should be taken. See Eq. (\ref{energy n=0}).}

\begin{eqnarray}
\nonumber
\xi_{m} & = & \sqrt{\left(s_{m}\lambda_{SO}-\eta_{m}elE_{z}\right)^{2}+\left(\hbar\omega_{L}\right)^{2}n_{m}}\label{sorted energies}\\
&  & -s_{m}\hbar\omega_{Z},
\end{eqnarray}
where $s_{m}=\pm1$ gives the spin, $n_{m}=0,1,2\ldots$ the LL and $\eta_{m}=\pm1$ the valley for the $m$ position. If we denote $\theta=q_{c}-q=N/D-\left[N/D\right]$ to
the occupancy factor of the last unfilled Landau level, the internal
energy at $T=0$ K is

\begin{equation}
	U=\sum_{m=0}^{q-1}D\xi_{m}+D\theta\xi_{q}.\label{UT}
\end{equation}
In Eq. (\ref{sorted energies}) we separate $\xi_{m}=\xi_{m}^{0}-s_{m}\hbar\omega_{Z}$,
with $\xi_{m}^{0}=\sqrt{\left(s_{m}\lambda_{SO}-\eta_{m}elE_{z}\right)^{2}+\left(\hbar\omega_{L}\right)^{2}n_{m}}$.
Replacing in Eq. (\ref{UT}) we can write

\begin{equation}
	U=U_{0}-D\hbar\omega_{Z}\left[\sum_{m=0}^{q-1}s_{m}+\theta s_{q}\right],\label{UT2}
\end{equation}
where

\begin{equation}
	U_{0}=\sum_{m=0}^{q-1}D\xi_{m}^{0}+D\theta\xi_{q}^{0}.\label{U0}
\end{equation}
The last term in Eq. (\ref{UT2}) can be related to the Pauli paramagnetism
associated with the spin population. This can be seen by considering
$N_{+}$ and $N_{-}$ total number of spin up and down, respectively.
For $q$ levels filled, let $k_{+}$ be the number of (+1) values
and $k_{-}$ the number of (-1) values in the sorting function $s_{m}$,
with $m=0,1,\ldots,q-1$ (thus $k_{+}$ and $k_{-}$ represents the
number of spin up and down states totally filled, respectively). Consequently,
$k_{+}+k_{-}=q$ and $\sum_{m=0}^{q-1}s_{m}=k_{+}-k_{-}$. For the
last unfilled level there could be two cases: (i) it is spin up or
(ii) spin down. For spin up, $s_{q}=1$ and therefore we can write
the total number of spin up and down as $N_{+}=Dk_{+}+D\theta s_{q}$
and $N_{-}=Dk_{-}$. Thus using that $\sum_{m=0}^{q-1}s_{m}=k_{+}-k_{-}$
we have $N_{+}-N_{-}=D\left[\sum_{m=0}^{q-1}s_{m}+\theta s_{q}\right]$.
The same result holds if the last unfilled level is spin down. Therefore
the Pauli magnetization is

\begin{equation}
	M_{P}=\mu_{B}\left(N_{+}-N_{-}\right)=\mu_{B}D\left[\sum_{m=0}^{q-1}s_{m}+\theta s_{q}\right].\label{Mp}
\end{equation}
Notice that this result is independent of how the energy levels are sorted, so that the last term in Eq. (\ref{UT2}) is always related
to the Pauli paramagnetism. Consequently, because $\hbar\omega_{Z}=\mu_{B}B$,
Eq. (\ref{UT2}) becomes

\begin{equation}
	U=U_{0}-BM_{P}.\label{UT3}
\end{equation}
The magnetization at $T=0$ K is $M=-\partial U/\partial B$. From Eq.
(\ref{UT3}) we have

\begin{equation}
	M=M_{0}+M_{P}+B\frac{\partial M_{P}}{\partial B},\label{M1}
\end{equation}
where $M_{0}=-\partial U_{0}/\partial B$. Given that $\partial D/\partial B=D/B$,
$\partial\theta/\partial B=-N/(DB)$ and $\partial\xi_{m}^{0}/\partial B=\left(\hbar\omega_{L}\right)^{2}n_{m}/2B\xi_{m}^{0}=\left[\xi_{m}^{0}-\left(s_{m}\lambda_{SO}-\eta_{m}elE_{z}\right)^{2}/\xi_{m}^{0}\right]/2B$,
we have

\begin{equation}
	M_{0}=\frac{1}{B}\left(N\xi_{q}^{0}-\frac{3}{2}U_{0}\right)+M',\label{M0}
\end{equation}
where 

\begin{eqnarray}
\nonumber
M' & = & \frac{D}{2B}\left[\sum_{m=0}^{q-1}\frac{\left(s_{m}\lambda_{SO}-\eta_{m}elE_{z}\right)^{2}}{\xi_{m}^{0}}\right.\label{M'}\\
&  & +\left.\frac{\left(s_{q}\lambda_{SO}-\eta_{q}elE_{z}\right)^{2}}{\xi_{q}^{0}}\right].
\end{eqnarray}
This new contribution $M'$ to the magnetization is not present in graphene
\cite{Escudero_2017} due to the negligible SOI and zero buckle height.
On the other hand, from Eq. (\ref{Mp}) we get $\partial M_{P}/\partial B=M_{P}/B+\mu_{B}Ds_{q}\partial\theta/\partial B=\left(M_{P}-\mu_{B}Ns_{q}\right)/B$.
Therefore, from Eqs. (\ref{sorted energies}), (\ref{UT3}) and (\ref{M0}),
the total ground state magnetization given by Eq. (\ref{M1}) can
be written as

\begin{equation}
	M=\frac{1}{B}\left(N\xi_{q}-\frac{3}{2}U\right)+M'+\frac{1}{2}M_{P}.\label{M2}
\end{equation}
This is the fundamental equation for our analysis. It shows that the
MO peaks are produced whenever $\xi_{q}$, $M'$ or $M_{P}$ changes
discontinuously, $U$ being continuous always. Thus the magnetization
in pristine silicene at $T=0$ K oscillates in a sawtooth pattern,
as in graphene \cite{Zhang_2010,Escudero_2017} and in general 2DEG with a Dirac-like
spectrum \cite{Sharapov_2004}. This is also in agreement with the results
found in \cite{Tabert_2015}, where the MO at $T=0$ K in a pristine
buckled honeycomb lattice are expressed as an infinite sum of harmonics
$k$ of the form $\sin(k)/k$,
which gives a sawtooth oscillation. From Eq. (\ref{M2}) we write
the MO peak amplitude $\Delta M$ as

\begin{equation}
	\Delta M=\frac{N}{B}\Delta\xi_{q}+\Delta M'+\frac{1}{2}\Delta M_{P}.\label{DeltaM-1}
\end{equation}
The first contribution $\Delta\xi_{q}$ comes directly from the discontinuous
change in the last energy level, which occurs only when the filling
factor $q$ changes. On the other hand, by analyzing Eqs. (\ref{Mp})
and (\ref{M'}) we see that the MO peaks produced by $\Delta M'$
and/or $\Delta M_{P}$ occur when the parameters $\eta_{q}$ and $s_{q}$
change but $\xi_{q}$ remains continuous. This would happen if the
change in $\eta_{q}$ and $s_{q}$ does not come from a change in the
filling factor $q$.

Eq. (\ref{M2}) also allows an intuitive interpretation of the effect
that impurities have in the magnetization. In the pristine case, the
discontinuities in $\xi_{q}$, $M'$ or $M_{P}$ are essentially a
product of the discrete LL, which gives a delta-like density of states
(DOS) and causes the MO to be sawtooth-like. But when impurities are added
to the system, the DOS is broaden and the discontinuities in $\xi_{q}$,
$M'$ or $M_{P}$ disappear. Consequently, the MO are also broaden and the
oscillations are no more sawtooth like.

\subsection{LL, valley and spin mixing}

\begin{figure}[t]
	\begin{center}
		\includegraphics[scale=0.5]{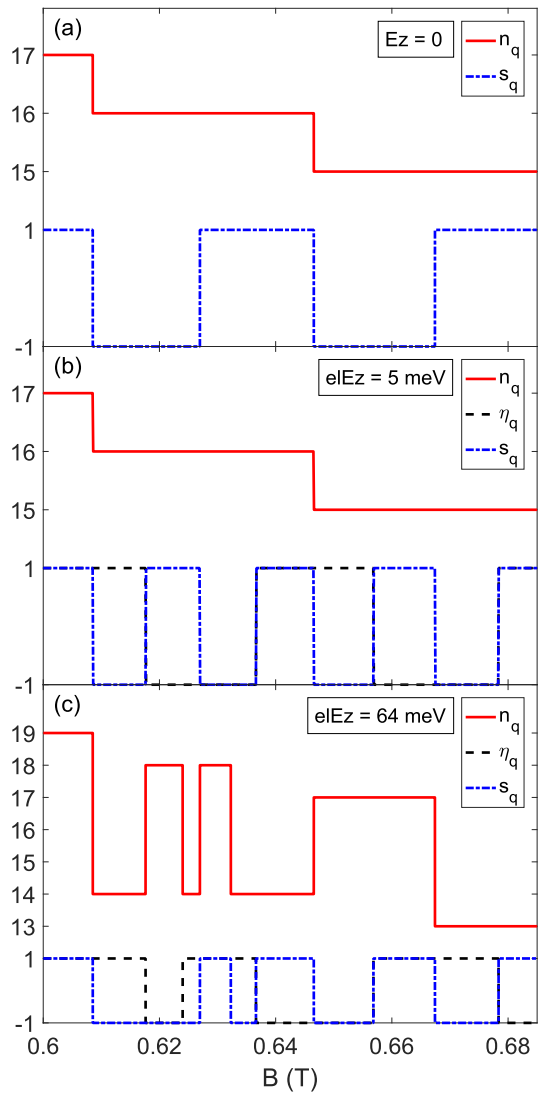}	
		\caption{\label{fig1}Parameters $n_{q}$ (LL), $\eta_{q}$ (valley) and $s_{q}$ (spin)
			as a function of the magnetic field $B$, for (a) $E_{z}=0$, (b)
			$elE_{z}=5$ meV and (c) $elE_{z}=64$ meV.}	
	\end{center}
\end{figure}

We are interested in how the parameters $n_{q}$, $\eta_{q}$ and
$s_{q}$ vary with $B$ for different values of $elE_{z}$. We recall
that $n_{q}$ may takes values $0,1,2,\ldots$, while $\eta_{m}=1\:(-1)$
for the $K\:(K')$ valley and $s_{m}=1\:(-1)$ for spin up (down).
The value of these parameters depends on the sorted position $q$,
which in turn depends in the mixing of the LL, valley and spin.
We consider a conduction electron density $n_{e}=0.01\:\mathrm{nm^{-2}}$
and an area $A=1000\:\mathrm{nm^{2}}$. In Fig. \ref{fig1} we show the parameters
$n_{q}$, $\eta_{q}$ and $s_{q}$ as a function of $B$ for different
values of $elE_{z}$. 

In Fig. \ref{fig1}(a) we can see the case $E_{z}=0$, where there is no parameter $\eta_q$  because each energy level in Eq. (\ref{sorted energies}) has
a doubly valley degeneracy. The last energy level $\xi_{q}$ changes
discontinuously only whenever the LL $n_{q}$, the spin $s_{q}$ or
both change. When $E_{z}\neq0$, the valley degeneracy is broken and the energy levels start to depend on $\eta_q$. This can be seen in Fig. \ref{fig1}(b), where for $elE_{z}=5$ meV the parameters start to vary differently. Nevertheless, it should be noted that in both Fig. \ref{fig1}(a) and \ref{fig1}(b) there is no appreciable mixing of the parameters. This means that in each case $n_q$ is always decreasing (as $B$ is increased), while $\eta_q$ and $s_q$ always alternate in the same way between -1 and 1. Moreover, in either case both $M'$ and $M_p$ are always continuous because every change in the parameters is produced by a change in the filling factor $q$, so $\Delta M'=0=\Delta M_p$ in Eq. (\ref{DeltaM-1}).
As $E_{z}$ increases,
the parameters start to vary in a more complicated way. This can be seen in Fig.
\ref{fig1}(c), in the case $elE_{z}=64$ meV, where there is a notorious mixing
of the parameters, which implies that $M'$ and $M_{p}$ may not be
always continuous. In such case all three contributions in Eq. (\ref{DeltaM-1})
will be present. 

\section{Classification of MO peaks}

We showed in Eq. (\ref{M2}) that the MO peaks are produced by the
discontinuous changes in $\xi_{q}$, $M'$ or $M_{P}$, which in turn
depends on $n_{q}$, $\eta_{q}$ and $s_{q}$. This allows a classification
of the MO peaks according to the parameters that change. We can define
seven general types of peaks, considering the change of LL, valley
or spin and its combinations. This can be seen in Table 1.

\begin{table}[h]\footnotesize
	\begin{center}	
		\begin{tabular}{c c c c}
			\hline
			LL change & Valley change & Spin change & Type of MO peak \\ \hline
			\checkmark & $\times$ & $\times$ & L \\ \hline
			\checkmark & \checkmark & $\times$ & LV \\ \hline
			\checkmark & $\times$ & \checkmark & LS \\ \hline
			\checkmark & \checkmark & \checkmark & LVS \\ \hline
			$\times$ & \checkmark & $\times$ & V \\ \hline
			$\times$ & \checkmark & \checkmark & VS \\ \hline
			$\times$ & $\times$ & \checkmark & S \\ 
			\hline 
		\end{tabular}
		
	\caption{Classification of MO peaks according to the change in 
		the parameters
		$n_{q}$ (LL), $\eta_{q}$ (valley) and $s_{q}$ (spin).}
\end{center}	
\end{table}

Furthermore, each type of MO peak has its own subpeaks, corresponding
to different possible ways in which the parameters can change. The
type of subpeak can be identified from the change in $n_{q}$, $\eta_{q}$
and $s_{q}$. In general we label the subpeaks as $\mathrm{X}_{\overline{\eta},\:\overline{s}}^{\overline{n}}$,
where $\mathrm{X=\left\{ L,\:LV,LS,\:LVS,V,\:VS,S\right\} }$ identifies
the type of MO peaks, as classify in Table 1, and $\overline{n}$,
$\overline{\eta}$ and $\overline{s}$ indicate the change (or not)
of the parameters. For example, consider an LS peak corresponding
to a change of LL $n=5\rightarrow4$ and spin $\uparrow\rightarrow\downarrow$, in the valley $K$. Then we identify
this peak with $\mathrm{LS_{K,\uparrow\rightarrow\downarrow}^{5\rightarrow4}}$.

The defined classification of MO peaks provides a systematic way of
recognizing them in a magnetization graph. One procedure could be
to first analyze how the energy levels are sorted (as done in Sec.
II.C), from which one could predict which type of MO peak appear and
in which order. We shall do this first for the case $elE_{z}\ll1$
eV, and then for the general case in which $elE_{z}$ may take any
value. In all cases we shall take $0.6 <B[T]<0.685$, as in Fig. \ref{fig1}.

\subsection{Limit $elE_{z}\ll1$ eV}

\begin{figure}[t]
	\includegraphics[scale=0.35]{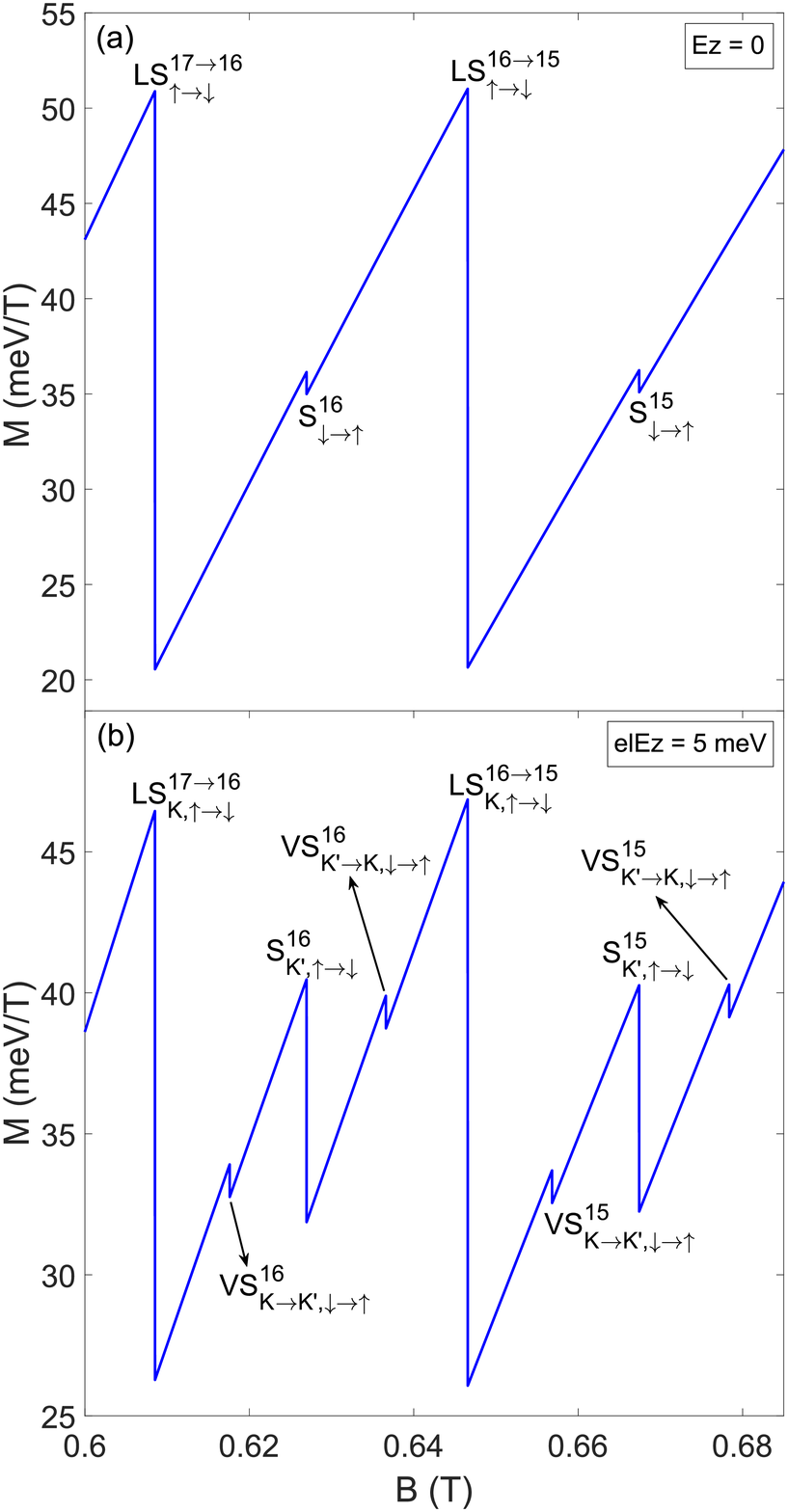}
	\caption{\label{fig2}Magnetization given by Eq. (\ref{M2}) for (a) $E_{z}=0$ and (b) $elE_{z}=5$
		meV.}
\end{figure}

We consider low $elE_{z}$ such that the parameters change only when the filling factor $q$ does it. Hence in this regime both $M'$ and $M_{p}$
are continuous and only $\Delta\xi_{q}$ contributes to the MO peaks
in Eq. (\ref{DeltaM-1}). We will consider the cases $E_{z}=0$ and $elE_{z}=5$ meV, when the sorting of the parameters is given by Fig. \ref{fig1}. Then,
following the classification of
Table 1, we see in Fig. \ref{fig1}(a) that the possible type of MO peaks at $E_{z}=0$ are LS and S, with the
order LS, S, LS, S,... On the other hand, we see in Fig. \ref{fig1}(b) that at $elE_{z}=5$ meV the MO peaks are LS, VS and S, with the order LS, VS, S, VS, LS... These results can be seen in Fig. \ref{fig2}, where we plot the
magnetization (\ref{M2}) for $E_{z}=0$ and $elE_{z}=5$ meV.

For $E_{z}=0$ we effectively see that only the peaks LS
and S appear.  
The S peaks always correspond to a change of spin down to up, so its amplitude is always $\Delta M_s=2N\mu_B$ at $E_z=0$.
When $elE_{z}=5$ meV, the VS peak appears,
and the order of the peaks is LS, VS, S,
VS, S,..., as expected. 
The LS peak always
correspond to a change of spin up to down in a $K$ valley. The VS
peak corresponds to $\left\{ K\rightarrow K',\downarrow\rightarrow\uparrow\right\}$
or $\left\{ K'\rightarrow K,\downarrow\rightarrow\uparrow\right\} ,$ while the S
peak always correspond to a change of spin up to down in the $K'$
valley. Thus, from Eqs. (\ref{sorted energies}) and (\ref{DeltaM-1})
we can write the corresponding peaks amplitude 

\begin{eqnarray}
\nonumber
\Delta M_{LS} & = & \frac{N}{B}\left[\chi\left(n_{q};\:-elE_{z}\right)-\chi\left(n_{q}-1;\:+elE_{z}\right)\right]\label{DeltaMLS}\\*
&  & -2N\mu_{B},\\
\Delta M_{VS} & = & 2N\mu_{B}\label{DeltaMVS}\\
\nonumber
\Delta M_{S} & = & \frac{N}{B}\left[\chi\left(n_{q};\:+elE_{z}\right)-\chi\left(n_{q};\:-elE_{z}\right)\right]\label{DeltaMS}\\*
&  & -2N\mu_{B},
\end{eqnarray}
where $\chi\left(n_{q};\:\pm elE_{z}\right)=\sqrt{\left(\lambda_{SO}\pm elE_{z}\right)^{2}+\left(\hbar\omega_{L}\right)^{2}n_{q}}$,
with $n_{q}$ being the corresponding LL level, as indicated in Fig.
\ref{fig2}. The VS peak always has the same amplitude $2N\mu_B$, which is equal to $\Delta M_s$ in the case $E_z=0$. 
In the limit $elE_{z}\ll1$ we can approximate the subpeaks amplitude
in Eqs. (\ref{DeltaMLS}) and (\ref{DeltaMS}) by

\begin{eqnarray}
\frac{B}{N}\Delta M_{LS} & \simeq & \varrho_{+}-2\hbar\omega_{Z}-\lambda_{SO}elE_{z}\varrho_{-},\label{DeltaMLS - low EZ}\\
\frac{B}{N}\Delta M_{S} & \simeq & \left[\gamma_{q}^{-1}\left(\lambda_{SO},\:\omega_{L}\right)2\lambda_{SO}\right]elE_{z}\label{DeltaMS - low EZ}\nonumber\\
&  & -2\hbar\omega_{Z}
\end{eqnarray}
where $\varrho_{\pm}=\gamma_{q}^{\pm1}\left(\lambda_{SO},\:\omega_{L}\right)\mp\gamma_{q-1}^{\pm1}\left(\lambda_{SO},\:\omega_{L}\right)$,
with $\gamma_{q}\left(\lambda_{SO},\:\omega_{L}\right)\equiv\sqrt{\lambda_{SO}^{2}+\left(\hbar\omega_{L}\right)^{2}n_{q}}$.
Thus the peaks amplitude is linear with $E_{z}$,
with both the slope and the y-intercept depending on $\lambda_{SO}$
and $\omega_{L}$. By studying how the amplitude of the peaks vary
with $E_{z}$, one could obtain the parameters
$\lambda_{SO}$ and $\omega_{L}$, provided that the magnetic field
$B$ and $n_{q}$ of the peaks are known. 
 It is important to notice that for each peak
the magnetic field $B$ and $n_{q}$ are different. The Landau level
$n_{q}$ could be inferred knowing at which $B$ the type of peak
occur, and how is the sorting of the energy levels. For instance, when $elE_{z}=5$ meV,
we know that the sorting is given by Fig. \ref{fig1}(b). Thus, the second peak LS in Fig. \ref{fig2}(b) corresponds to a change of LL from $n=16\rightarrow15$,
so we put $n_{q}=16$ in Eq. (\ref{DeltaMLS - low EZ}).

This way of obtaining $\lambda_{SO}$ and $\omega_{L}$ from the MO
peaks could be an useful alternative to the other available methods.
The Landau energy $\omega_{L}$ gives the Fermi velocity, since we
define $\omega_{L}=\upsilon_{F}\sqrt{2eB/\hbar}$. In silicene,
the Fermi velocity has usually been obtained using DFT or TB models,
with the result of a lower value than in graphene \cite{Guzm_n_Verri_2007,Cahangirov_2009,Dzade_2010}.
This can be easily understood from the reduced hopping in silicene
since the Si atoms are more distant from each other. Likewise, the
SO parameter $\lambda_{SO}$ is usually obtained from TB models using
the hopping parameters \cite{Liu_2011}.

\subsection{General case}

\begin{figure}[t]
	\includegraphics[scale=0.35]{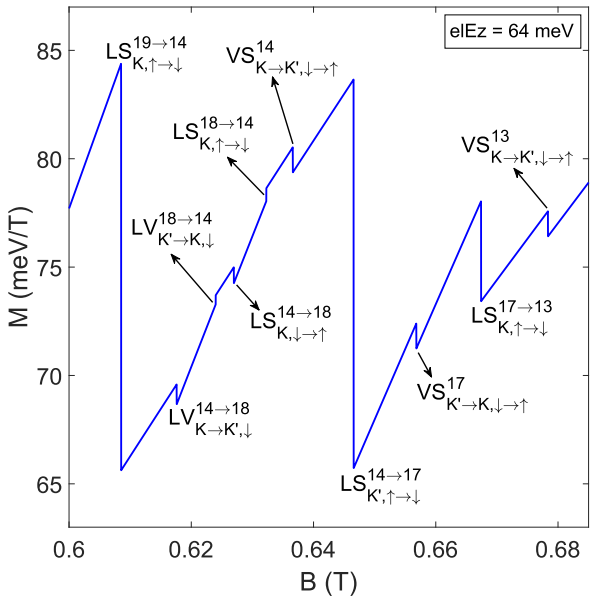}
	\caption{\label{fig3}Magnetization given by Eq. (\ref{M2}) for $elE_{z}=64$ meV.}	
\end{figure}

\begin{figure}[t]
	\includegraphics[scale=0.5]{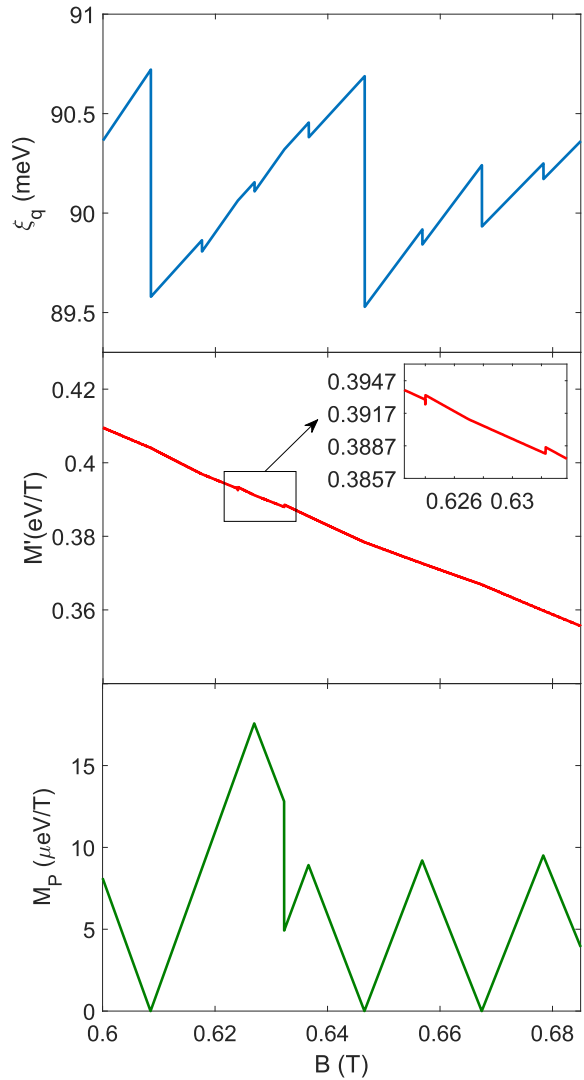}
	\caption{\label{fig4}$\xi_{q}$, $M'$ and $M_{P}$ given by Eqs. (\ref{sorted energies}),
		(\ref{Mp}) and (\ref{M'}), as a function of $B$, in the case $elE_{z}=64$
		meV. }
\end{figure}

In the general case the mixing of the parameters $n_{q}$, $\eta_{q}$
and $s_{q}$ depends on the specific value of $elE_{z}$. We saw in
Fig. \ref{fig1}(c) that these parameters vary in a complicated way as $elE_{z}$
increases. Thus for any specific value of electric field one should
see how the energy levels are sorted in order to identify the MO peaks.
Moreover, in the general case $M'$ and $M_{P}$ may no longer be
continuous, and the contributions $\Delta M'$ and $\Delta M_{p}$
should be taken into account in Eq. (\ref{DeltaM-1}). 

We consider $elE_{z}=64$ meV, in which case $n_{q}$, $\eta_{q}$
and $s_{q}$ as a function of $B$ are given in Fig. \ref{fig1}(c). Then we can
identify ten MO peaks for $0.6<B(T)<0.685$. For instance, the
first is a LS peak corresponding to $\left\{ K,\uparrow\rightarrow\downarrow\right\}$,
while the sixth is a VS peak corresponding to $\left\{ K\rightarrow K',\downarrow\rightarrow\uparrow\right\}$.
In the same way one can classify the other peaks, leading to the magnetization
for $elE_{z}=64$ meV shown in Fig. \ref{fig3}.

As we see, the mixing of the parameters alters drastically the magnetization.
In particular we note that in the two peaks $\mathrm{LV_{K'\rightarrow K,\downarrow}^{18\rightarrow14}}$
and $\mathrm{LS_{K,\uparrow\rightarrow\downarrow}^{18\rightarrow14}}$ the
magnetization increases, which is opposite to all other peaks, where
the magnetization always decreases. This feature suggests that the
peaks $\mathrm{LV_{K'\rightarrow K,\downarrow}^{18\rightarrow14}}$
and $\mathrm{LS_{K,\uparrow\rightarrow\downarrow}^{18\rightarrow14}}$ are
not produced by the change $\Delta\xi_{q}$, but come from the discontinuities
in $M'$ and $M_{p}$. Indeed, the contribution $\Delta\xi_{q}$ always
lower the magnetization because it comes from the discontinuous change
in the last energy level $\xi_{q}$ as $B$ increases. To see this
we plot in Fig. \ref{fig4} the variation of $\xi_{q}$, $M'$ and $M_{P}$ with
respect to $B$, for $elE_{z}=64$ meV.

We can clearly appreciate two discontinuities in $M'$ and one
discontinuity in $M_{P}$. We also see that when this discontinuities
occur $\xi_{q}$ is continuous\footnote{Nevertheless we can see that the slope of $\xi_{q}$ slightly changes when
$M'$ or $M_{P}$ are discontinuous. The reason for this is that the
variation of the parameters in this places do modify $\xi_{q}$, but
in a continuous way, giving a different dependence with $B$ without
any discontinuous jump.}, which implies $\Delta\xi_{q}=0$. Thus the peaks $\mathrm{LV_{K'\rightarrow K,\downarrow}^{18\rightarrow14}}$
and $\mathrm{LS_{K,\uparrow\rightarrow\downarrow}^{18\rightarrow14}}$ are
effectively produced by $\Delta M'$ and $\Delta M_{p}$. The first
peak $\mathrm{LV_{K'\rightarrow K,\downarrow}^{18\rightarrow14}}$ only has contribution from $\Delta M'$ because only the valley changes. This can be seen in Fig. \ref{fig4}, where when the first
discontinuity occurs in $M'$ we see that $M_{P}$ is continuous. On the other hand, the peak $\mathrm{LS_{K,\uparrow\rightarrow\downarrow}^{18\rightarrow14}}$
has both contributions $\Delta M'$ and $\Delta M_{p}$, as can be
seen in Fig. \ref{fig4}, where both $M'$ and $M_{P}$ have a discontinuity.
In this way we can say, in general, that the MO peaks that increase
the magnetization are not produced by the discontinuous change in
the last energy level $\xi_{q}$, but rather by the discontinuous
change in $M'$ and/or $M_{P}$.

\section{MO frequencies}

\begin{figure}[t]
	\includegraphics[scale=0.5]{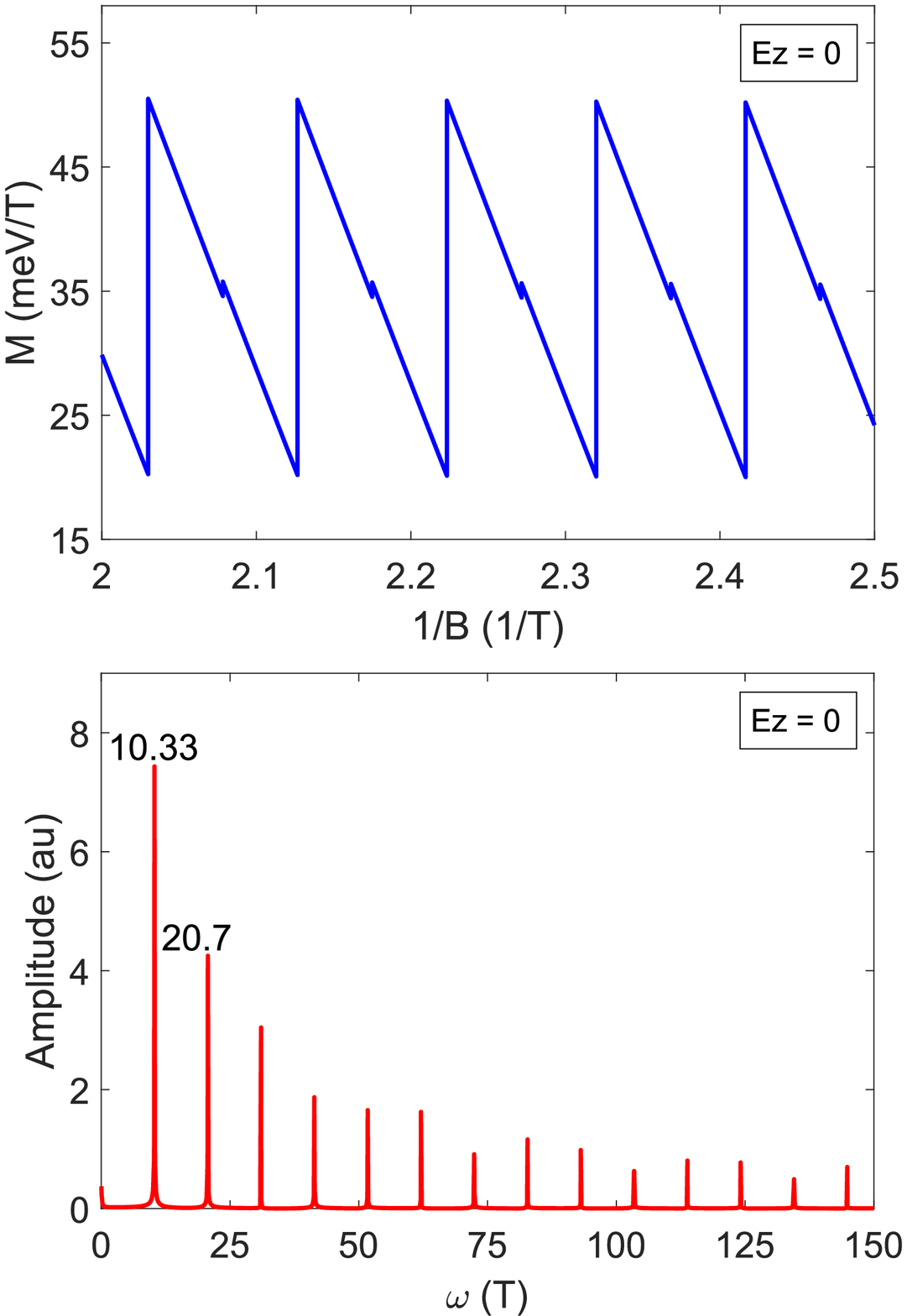}
	\caption{\label{fig6}Magnetization and fast Fourier transform (FFT) for $E_{z}=0$.}
\end{figure}

\begin{figure}[t]
	\includegraphics[scale=0.5]{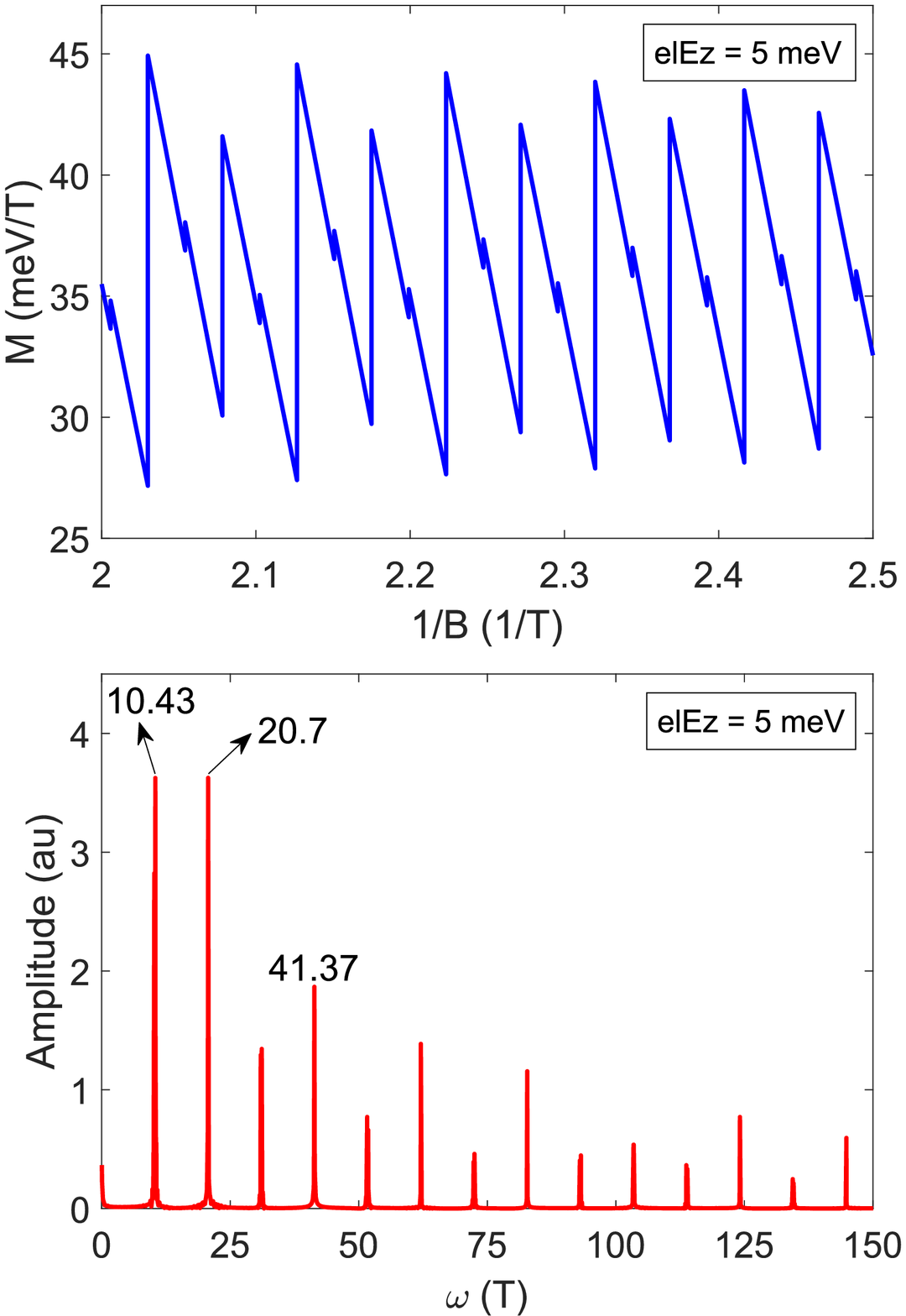}
	\caption{\label{fig7}Magnetization and fast Fourier transform (FFT) for $elE_{z}=5$ meV.}
\end{figure}

So far we have analyzed only the MO peaks amplitude, but we can also
obtain information from their frequencies. For simplicity we shall consider the cases $E_{z}=0$ and $elE_{z}=5$ meV, such that the only contribution in
Eq. (\ref{DeltaM-1}) is given by $\Delta\xi_{q}$ and all the MO peaks
can occur only when the filling factor $q$ changes. Given that
$q=\left[N/D\right]=\left[n_{e}\varphi/B\right]$, it is clear that
the magnetization oscillates periodically as a function of $1/B$,
in agreement with the Onsager relation \cite{Onsager_1952}. The period of oscillation
is in general given by $\Delta\left(1/B\right)=1/B_{2}-1/B_{1}$, where $B_{1}=n_{e}\varphi/q_{1}$
and $B_{2}=n_{e}\varphi/q_{2}$ ($\varphi=h/e$). Thus we can write

\begin{equation}
\Delta\left(\frac{1}{B}\right)=\frac{e}{2\pi\hbar n_{e}}\Delta q,\label{period}
\end{equation}
where $\Delta q=q_{2}-q_{1}$. Because the MO are sawtooth like, there
will be many frequencies involved in its Fourier expansion. Nevertheless
we are interested only in the fundamental frequencies, for the others
are just harmonics of these ones. To obtain the frequency spectrum
we performed a fast Fourier transform (FFT) in the magnetization as
a function of $1/B$. 

In Fig. \ref{fig6} we can appreciate the case $E_{z}=0$, where the MO are a combination
of two sawtooth oscillations (SO) with different frequencies, as can be inferred
in the FFT spectrum, where two main frequencies $\omega_{1}=10.33$
T and $\omega_{2}=20.67$ T can be recognized. This can be explained
if we decompose the term $\xi_{q}$ in Eq. (\ref{M2}), which causes
the SO with its discontinuous change. When $E_{z}=0$ we have, from
Eq. (\ref{sorted energies}), $\xi_{q}=\sqrt{\lambda_{SO}^{2}+\left(\hbar\omega_{L}\right)^{2}n_{q}}-s_{q}\hbar\omega_{Z}=\xi_{q}^{L}-\xi_{q}^{S}$,
where we separated

\begin{eqnarray}
\xi_{q}^{L} & = & \sqrt{\lambda_{SO}^{2}+\left(\hbar\omega_{L}\right)^{2}n_{q}},\label{Esorted L}\\
\xi_{q}^{S} & = & s_{q}\hbar\omega_{Z}\label{Esorted S}.
\end{eqnarray}
The term $\xi_{q}^{L}$ is only related to the LL, while the term $\xi_{q}^{S}$ is only related to the spin. 
Then, considering the parameters sorting given by Fig. \ref{fig1}(a), and taking into account the valley degeneracy at $E_z=0$, we get that $\xi_{q}^{L}$ changes periodically
when $q$ changes by four, so $\Delta q=4$ in Eq. (\ref{period}),
giving the frequency $\omega_{1}=\pi\hbar n_{e}/2e$. On the other
hand, $\xi_{q}^{S}$ changes
periodically when $q$ changes by two, so $\Delta q=2$ in Eq.
(\ref{period}), giving the frequency $\omega_{2}=\pi\hbar n_{e}/e$.
Thus the magnetization in Eq. (\ref{M2}) can be decomposed in two
SO with two different fundamental frequencies. For the
considered electron density $n_{e}=0.01\:\mathrm{nm^{-2}}$ we obtain
$\omega_{1}=10.34$ T and $\omega_{2}=20.68$ T, in agreement with
Fig. \ref{fig6}.

When $elE_{z}=5$ meV the valley degeneracy is broken, which  gives rise to another MO frequency $\omega_{3}=41.33$ T, as seen in
Fig. \ref{fig7}. 
The origin of this can be explained by first approximating
$\xi_{q}$ for low $elE_{z}$, so $\xi_{q}\simeq\sqrt{\lambda_{SO}^{2}+\left(\hbar\omega_{L}\right)^{2}n_{q}}-s_{q}\hbar\omega_{Z}-elE_{z}\lambda_{SO}s_q\eta_{q}\left[\lambda_{SO}^{2}+\left(\hbar\omega_{L}\right)^{2}n_{q}\right]^{-1/2}$.
This can be separated as $\xi_{q}=\xi_{q}^{L}-\xi_{q}^{S}-\xi_{q}^{VS}$,
where $\xi_{q}^{L}$ and $\xi_{q}^{S}$ are given by Eqs. (\ref{Esorted L}) and (\ref{Esorted S}), while

\begin{equation}
\xi_{q}^{VS} = elE_{z}\frac{\lambda_{SO}\eta_{q}s_{q}}{\sqrt{\lambda_{SO}^{2}+\left(\hbar\omega_{L}\right)^{2}n_{q}}}.\label{Esorted VS}
\end{equation}
For $elE_{z}=5$ meV, the sorting of the parameters $n_{q}$,
$\eta_{q}$ and $s_{q}$ is given by Fig. \ref{fig1}(b). Hence $n_{q}$ still changes only $q$ changes by four, so
$\xi_{q}^{L}$ gives the frequency $\omega_{1}=\pi\hbar n_{e}/2e=10.34$ T. 
On the other hand, now $s_q$ changes whenever $q$ changes, so for $\xi_{q}^{S}$ we have $\Delta q=1$ in Eq. (\ref{period}). This gives a new frequency $\omega_{3}=2\pi\hbar n_{e}/e$ which implies
$\omega_{3}=41.36$ T for $n_{e}=0.01\:\mathrm{nm^{-2}}$. Finally, the new defined term $\xi_{q}^{VS}$ in Eq. (\ref{Esorted VS}) changes discontinuously when $\Delta q=2$, as can be easily seen in Fig. \ref{fig1}(b). Therefore we also have the frequency $\omega_{2}=\pi\hbar n_{e}/e=20.68$ T.

\section{Conclusions}
We studied the magnetic oscillations (MO) in pristine silicene at
$T=0$ K. We considered a constant electron density, such that the
valence band is full and only the conduction band is available. Under
a perpendicular electric and magnetic field, we found analytical expressions
for the ground state internal energy and magnetization. We obtained
that the MO are sawtooth-like and are entirely produced by the change
in the last energy level occupied. This lead us to a classification
of the MO peaks in terms of the parameters $n_{q}$ (LL), $\eta_{q}$
(valley) and $s_{q}$ (spin) which define the last energy level. In
general we defined seven types of MO peaks, as indicated in Table
1. Using this classification we analyzed the MO in the case of low
electric field ($elE_{z}\ll1$ eV), and the general case in which
$E_{z}$ may take any value. In each case we were able to classify
the type of MO present, and in which order. When $E_{z}=0$ the energy
levels have a valley degeneracy and the MO peaks occur only when the
last occupied level changes its LL and/or spin. On the other hand,
when $E_{z}\neq0$ the valley degeneracy is broken and new MO peaks
appear, associated with the change of valley in the last energy level.
Furthermore, we found that analyzing the MO peak amplitude one could
extract information about the Fermi velocity and the spin-orbit interaction
strength, which could be an useful alternative to the other available
methods. For the general case of $E_{z}$ the last energy level varies
in a complicated way and therefore so does it the MO peaks. Nevertheless
one can still classify the peaks by studying the change in the parameters
$n_{q}$, $\eta_{q}$ and $s_{q}$ at any particular $E_{z}$. Finally we analyzed the
MO frequencies, where we found that the magnetization effectively oscillates
periodically a function of $1/B$. We performed the fast Fourier transform
spectrum of the sawtooth-like MO oscillations. When $E_{z}=0$ we
found two fundamental frequencies, corresponding to the change of
LL or spin in the last energy level. When $elE_{z}=5$ meV a new
fundamental frequency appears, associated with broken valley degeneracy.

\section{Acknowledgment}
This paper was partially supported by grants of CONICET (Argentina
National Research Council) and Universidad Nacional del Sur (UNS)
and by ANPCyT through PICT 2014-1351. Res. N 270/15. N: 2014-1351, and PIP 2014-2016. Res. N 5013/14. Código: 11220130100436CO research grant, as well as by SGCyT-UNS., J.
S. A. and P. J. are members of CONICET., F. E. acknowledge  research fellowship from this institution.

\appendix
\section{Magnetization for constant Fermi energy}

We shall analyze the case in which the Fermi energy $\mu$ is held
constant, instead of the conduction electron density $n_{e}$. We
consider $\mu>0$ such that last energy level filled always correspond
to the CB. The valence band is not taken into account since it is full and thus will
not contribute to the MO. Because we consider $\mu$ fixed, whereas
the number of electrons $N$ may change, we work with the grand potential
$\Omega$. For a Fermi energy $\mu$, all energies levels $m=0,1,2\ldots,m_{F}$
all filled, where $m_{F}$ is such that $\xi_{m_{F}}\leq\mu\leq\xi_{m_{F+1}}$.
Then the grand potential $\Omega$ at $T=0$ K is

\begin{equation}
\Omega=\sum_{m=0}^{m_{F}}D\left(\xi_{m}-\mu\right),\label{GP}
\end{equation}
where $\xi_{m}$ is given by Eq. (\ref{sorted energies}). Separating
$\xi_{m}=\xi_{m}^{0}-s_{m}\hbar\omega_{Z}$, with $\xi_{m}^{0}=\sqrt{\left(s_{m}\lambda_{SO}-\eta_{m}elE_{z}\right)^{2}+\left(\hbar\omega_{L}\right)^{2}n_{m}}$,
we get

\begin{equation}
\Omega=\Omega_{0}-D\hbar\omega_{Z}\sum_{m=0}^{q-1}s_{m},\label{GP2}
\end{equation}
where

\begin{equation}
\Omega_{0}=\sum_{m=0}^{m_{F}}D\left(\xi_{m}^{0}-\mu\right).\label{GP0}
\end{equation}
As in the case with $N$ constant, the last term in Eq. (\ref{GP2})
is related to the Pauli paramagnetism associated with the spin population,
with the difference that this time all the energy levels below the
Fermi energy are completely filled. Thus the Pauli paramagnetism is
$M_{P}=\mu_{B}\left(N_{+}-N_{-}\right)=\mu_{B}D\sum_{m=0}^{m_{F}}s_{m}$,
and Eq. (\ref{GP2}) becomes

\begin{equation}
\Omega=\Omega_{0}-BM_{P}.\label{GP3}
\end{equation}
This result is similar to the one obtained in Eq. (\ref{UT3}), with
$U$ being replaced by $\Omega$. Therefore, similar expressions are
obtained for the magnetization, given by $M=-\left(\partial\Omega/\partial B\right)_{\mu}$.
From Eq. (\ref{GP3}) we obtain

\begin{equation}
M=-\frac{1}{2B}\left(3\Omega+N\mu\right)+M'+\frac{1}{2}M_{P},\label{M2-1}
\end{equation}
where $N=\sum_{m=0}^{m_{F}}D=D\left(m_{F}-1\right)$ is the number
of electrons, and

\begin{equation}
M'=\frac{D}{2B}\sum_{m=0}^{m_{F}}\frac{\left(s_{m}\lambda_{SO}-\eta_{m}elE_{z}\right)^{2}}{\xi_{m}^{0}}.
\end{equation}
Eq. (\ref{M2-1}) shows that, when $\mu$ is constant, the MO peaks
are produced whenever $N$, $M'$ or $M_{P}$ changes discontinuously,
with $\Omega$ being continuous always. Then in general we write the MO
peak amplitude $\Delta M$ as

\begin{equation}
\Delta M=-\frac{\mu}{2B}\Delta N+\Delta M'+\frac{1}{2}\Delta M_{P}.\label{DeltaM-1-1}
\end{equation}
This last equation is similar to Eq. (\ref{DeltaM-1}), with each
contribution $\Delta N$, $\Delta M'$ and $\Delta M_{P}$ being still
defined by the discontinuous change in the parameters $n_{q}$, $\eta_{q}$
and $s_{q}$. The main difference is that in this case, with $\mu$
constant, all the three functions $N$, $M'$ and $M_{P}$ have discontinuities
at any $E_{z}$. Nevertheless, one could still classify the MO peaks
as done in Table 1, which accounts for the main results found in the
case when $N$ is constant.

\section*{References}

\bibliography{bibliography}	

\end{document}